\documentclass[twocolumn,superscriptaddress,aps,pra,preprintnumbers,amsmath,amssymb,floatfix,groupedaddress]{revtex4-1}
\usepackage[T1]{fontenc}
\usepackage[utf8]{inputenc}
\usepackage{float}
\usepackage{amsmath}
\usepackage{amssymb}
\usepackage{graphicx}
\usepackage{esint}

\makeatletter

\@ifundefined{textcolor}{}
{%
 \definecolor{BLACK}{gray}{0}
 \definecolor{WHITE}{gray}{1}
 \definecolor{RED}{rgb}{1,0,0}
 \definecolor{GREEN}{rgb}{0,1,0}
 \definecolor{BLUE}{rgb}{0,0,1}
 \definecolor{CYAN}{cmyk}{1,0,0,0}
 \definecolor{MAGENTA}{cmyk}{0,1,0,0}
 \definecolor{YELLOW}{cmyk}{0,0,1,0}
}

\makeatother

\begin{document}

\title{Quantum optical non-linearities induced by Rydberg-Rydberg interactions
: \\
 a perturbative approach}

\author{A. Grankin$^{1}$, E. Brion$^{2}$, E. Bimbard$^{1}$, R. Boddeda$^{1}$,
I. Usmani$^{1}$, A. Ourjoumtsev$^{1}$, P. Grangier$^{1}$ }

\affiliation{$^{1}$Laboratoire Charles Fabry, Institut d'Optique, CNRS, Univ.
Paris-Sud, 2 Avenue Fresnel, 91127 Palaiseau, France \\
 $^{2}$Laboratoire Aimé Cotton, CNRS, Université Paris Sud, ENS Cachan,
91405 Orsay, France.}
\begin{abstract}
In this article, we theoretically study the quantum statistical properties
of the light transmitted through or reflected from an optical cavity,
filled by an atomic medium with strong optical non-linearity induced
by Rydberg-Rydberg van der Waals interactions. Atoms are driven on
a two-photon transition from their ground state to a Rydberg level
via an intermediate state by the combination of a weak signal field
and a strong control beam. By using a perturbative approach, we get
analytic results which remain valid in the regime of weak feeding
fields, even when the intermediate state becomes resonant. Therefore
they allow us to investigate quantitatively new features associated
with the resonant behaviour of the system. We also propose an effective
non-linear three-boson model of the system which, in addition to leading
to the same analytic results as the original problem, sheds light
on the physical processes at work in the system. 
\end{abstract}

\pacs{32.80.Ee, 42.50.Ar, 42.50.Gy, 42.50.Nn}

\maketitle

\section{Introduction}

Photons are considered as the best long-range quantum information
carriers; they, however, do not directly interact with each other,
which makes the processing of the information they carry problematic
\cite{CVL14}. Standard Kerr dispersive non linearities obtained in
non-interacting atomic ensembles, either in off-resonant two-level
or resonant three-level configurations involving Electromagnetically
Induced Transparency (EIT), are usually too small to allow for quantum
non-linear optical manipulations. Among other techniques \cite{CVL14},
a possible way to enhance the non-linear susceptibility is to resort
to a Rydberg level as one of the long-lived states involved in the
EIT process \cite{PMG10,DK12,PFLHG12,MSB12,GBB14} : in such Rydberg
EIT protocols, the strong van der Waals interactions between Rydberg
atoms create a cooperative Rydberg blockade phenomenon \cite{LCFD01,SWM10,CP10},
where each Rydberg atom prevents the excitation of its neighbors inside
a \textquotedbl{}blockade sphere\textquotedbl{} and deeply changes
the EIT profile. In particular, giant dispersive non-linear effects
were experimentally obtained in an off-resonant Rydberg-EIT scheme
using cold rubidium atoms placed in an optical cavity \cite{PBS12,SPB13}.
In a previous paper \cite{GBB14}, we theoretically investigated the
quantum statistical properties of the light generated by this scheme
in the dispersive regime, \emph{i.e.} for strongly detuned intermediate
state. We showed that, under some assumptions, the system effectively
behaves as a large spin coupled to the cavity mode \cite{GBEM10}
and we computed the steady-state second-order correlation function
to characterize the bunched or antibunched emission of photons out
of the cavity.

In the present paper, we deal with the same system, but in a different
approach. Restricting ourselves to the low feeding regime, we present
an analytic derivation of the correlation function $g^{\left(2\right)}\left(\tau\right)$
for the transmitted and reflected light, based on the factorization
of the lowest perturbative order of operator product averages. It
is important to note that this derivation is valid in both the dispersive
and resonant regimes and therefore generalizes our previous results.
This factorization property is demonstrated rigorously for purely
radiative damping, but we show also that it is approximately preserved
in the experimentally relevant case of additional dephasing due to,
e.g., laser frequency and intensity noise. In addition, we propose
an effective non-linear three-boson model for the coupled atom-cavity
system which allows us to obtain the same results as the (more cumbersome)
exhaustive treatment. In the dispersive regime, this Hamiltonian agrees
with the one we obtained in the so-called ``Rydberg-bubble approximation''
\cite{GBB14}; it also allows us to investigate the dissipation at
work in the resonant case.

The paper is structured as follows. In Sec. \ref{The system}, we
recall our setup and the assumptions we make to compute its dynamics.
In Sec. \ref{g2}, we present an analytical way to obtain the correlation
functions for the light outgoing from the cavity and discuss some
of the numerical results we obtained. In Sec. \ref{ModelHam}, we
present and discuss an effective three-boson model, allowing us to
recover and generalize the previous results. Finally, we conclude
in Sec. \ref{sec:Conclusion} by evoking open questions and perspectives
of our work. Appendices address supplementary technical details which
are omitted in the text for readability.

\section{The system \label{The system}}

\begin{figure}
\begin{centering}
\includegraphics[width=8.2cm]{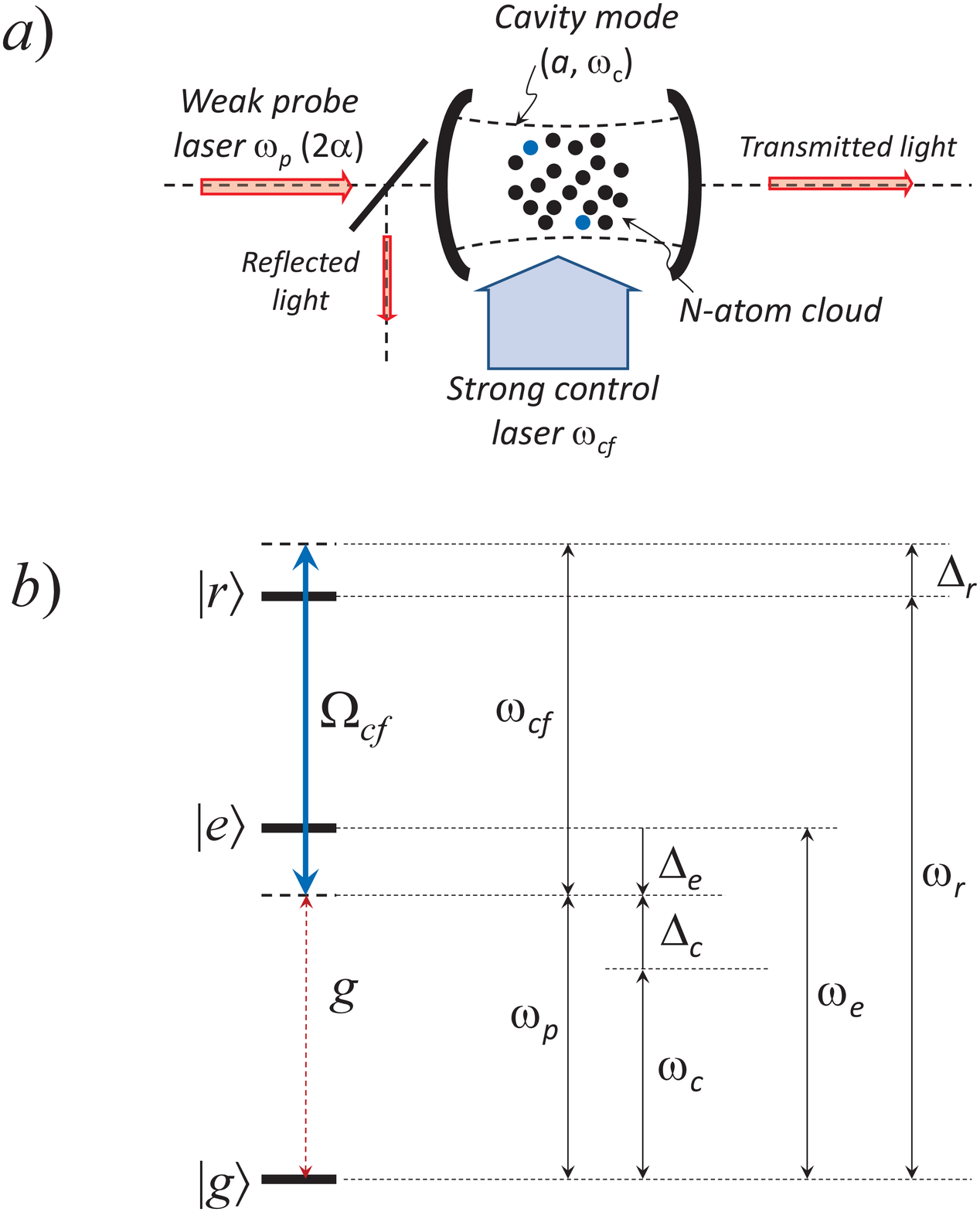} 
\par\end{centering}

\protect\caption{a) The setup consists of $N$ cold atoms placed in an optical cavity
which is fed by a weak (classical) laser beam of frequency $\omega_{p}$
and a strong control laser field of frequency $\omega_{cf}$. b) The
atoms present a three-level ladder structure $\left\{ \left|g\right\rangle ,\left|e\right\rangle ,\left|r\right\rangle \right\} $.
The transitions $\left|g\right\rangle \leftrightarrow\left|e\right\rangle $
and $\left|e\right\rangle \leftrightarrow\left|r\right\rangle $ are
driven by the injected probe and control laser fields, respectively,
with the respective coupling strength and Rabi frequency $g$ and
$\Omega_{cf}$ (see the text for the definitions of the different
detunings represented here).}

\label{FigSys} 
\end{figure}

The system we consider here is the same we dealt with in \cite{GBB14}.
It comprises $N$ atoms which present a three-level ladder structure
with a\emph{ }ground $\left|g\right\rangle $, intermediate $\left|e\right\rangle $
and Rydberg states $\left|r\right\rangle $ (see Figure \ref{FigSys}).
The energy of the atomic level $\left|k=g,e,r\right\rangle $ is denoted
by $\hbar\omega_{k}$ (by convention $\omega_{g}=0$) and the dipole
decay rates are $\gamma_{e}$ (intermediate state) and $\gamma_{r}$
(Rydberg state). The transitions $\left|g\right\rangle \leftrightarrow\left|e\right\rangle $
and $\left|e\right\rangle \leftrightarrow\left|r\right\rangle $ are
respectively driven by a weak probe field of frequency $\omega_{p}$
and a strong control field of frequency $\omega_{cf}$. Both fields
can \emph{a priori} be resonant or not with atomic transitions, the
respective detunings being defined by $\Delta_{e}\equiv\left(\omega_{p}-\omega_{e}\right)$
and $\Delta_{r}\equiv\left(\omega_{p}+\omega_{cf}-\omega_{r}\right)$.
Moreover, the atoms are placed in an optical cavity: we shall denote
by $\gamma_{c}^{\left(L,R\right)}$ the respective decay rates through
the left and right mirrors (see Fig. \ref{FigSys}), with $\gamma_{c}\equiv\gamma_{c}^{\left(L\right)}+\gamma_{c}^{\left(R\right)}$.
The transition $\left|g\right\rangle \leftrightarrow\left|e\right\rangle $
is supposed in the neighborhood of a cavity resonance. The frequency
and annihilation operator of the corresponding mode are denoted by
$\omega_{c}$ and $a$, respectively ; the detuning of this mode with
the probe laser is defined by $\Delta_{c}\equiv\left(\omega_{p}-\omega_{c}\right)$
and $\alpha$ denotes the feeding rate of the cavity mode with the
probe field, which is supposed real for simplicity. Finally, we introduce
$g$ and $\Omega_{cf}$ which are the single-atom coupling constant
of the transition $\left|g\right\rangle \leftrightarrow\left|e\right\rangle $
with the cavity mode and the Rabi frequency of the control field on
the transition $\left|e\right\rangle \leftrightarrow\left|r\right\rangle $,
respectively. As represented on Figure \ref{FigSys}, the setup allows
one to measure the statistics of both the reflected and transmitted
lights, i.e. $g^{\left(2\right)}\left(\tau\right)$.\textbf{ }

The dynamics of the full system, including the bath modes, are governed
by the Hamiltonian derived in Appendix \ref{Hamiltonian}, in the
Rotating Wave Approximation. We note that this Hamiltonian description
does not take into account any additional dephasing due to, e.g.,
laser intensity or frequency fluctuations : decays and dephasing are
therefore purely radiative. In the Markov approximation, the corresponding
Heisenberg-Langevin equations are 
\begin{widetext}
\begin{eqnarray}
 &  & \frac{d}{dt}\; a\;=\mathrm{i}D_{c}a-\mathrm{i}\alpha-\mathrm{i}g\sum_{i}^{N}\sigma_{ge}^{\left(i\right)}+\sqrt{2\gamma_{c}^{\left(L\right)}}a_{in}^{\left(L\right)}+\sqrt{2\gamma_{c}^{\left(R\right)}}a_{in}^{\left(R\right)}\label{HL1}\\
 &  & \frac{d}{dt}\sigma_{ge}^{\left(i\right)}=\mathrm{i}D_{e}\sigma_{ge}^{\left(i\right)}-\mathrm{i}\frac{\Omega_{cf}}{2}\sigma_{gr}^{\left(i\right)}+\mathrm{i}ga\left(\sigma_{ee}^{\left(i\right)}-\sigma_{gg}^{\left(i\right)}\right)+F_{ge}^{\left(i\right)}\label{HL2}\\
 &  & \frac{d}{dt}\sigma_{gr}^{\left(i\right)}=\mathrm{i}D_{r}\sigma_{gr}^{\left(i\right)}-\mathrm{i}\frac{\Omega_{cf}}{2}\sigma_{ge}^{\left(i\right)}+\mathrm{i}ga\sigma_{er}^{\left(i\right)}-\mathrm{i}\sigma_{gr}^{\left(i\right)}\sum_{j\neq i}^{N}\kappa_{ij}\sigma_{rr}^{\left(j\right)}+F_{gr}^{\left(i\right)}\label{HL3}\\
 &  & \frac{d}{dt}\sigma_{er}^{\left(i\right)}=\mathrm{i}D_{er}\sigma_{er}^{\left(i\right)}+\mathrm{i}\frac{\Omega_{cf}}{2}\left(\sigma_{rr}^{\left(i\right)}-\sigma_{ee}^{\left(i\right)}\right)+\mathrm{i}ga^{\dagger}\sigma_{gr}^{\left(i\right)}-\mathrm{i}\sigma_{er}^{\left(i\right)}\sum_{j\neq i}^{N}\kappa_{ij}\sigma_{rr}^{\left(j\right)}+F_{er}^{\left(i\right)}\label{HL4}
\end{eqnarray}
where $a_{in}^{\left(L\right)}$, $a_{in}^{\left(R\right)}$ and $F_{\alpha\beta}^{\left(i\right)}$
denote Langevin forces associated to the incoming fields from left
and right sides and to the atomic operator $\sigma_{\alpha\beta}^{\left(i\right)}$,
respectively. We also introduced the complex effective detunings $D_{k}\equiv\left(\Delta_{k}+\mathrm{i}\gamma_{k}\right)$
for $k=c,e,r$ and $D_{er}\equiv\left(\Delta_{r}-\Delta_{e}\right)+\mathrm{i}\left(\gamma_{r}+\gamma_{e}\right)$.
Note that we chose to make the feeding factor $\alpha$ appear explicitly
in Eq.(\ref{HL1}): in technical terms, it corresponds to displacing
the incoming field from the coherent state $\left|\alpha\right\rangle $
to the vacuum $\left|0\right\rangle $ ; to be consistent with this
choice, from now on, we must set $\left\langle a_{in}\right\rangle =0$.
\end{widetext}

In the next section, we show how to compute the correlation function
$g^{\left(2\right)}\left(\tau\right)$ at the lowest order in the
feeding parameter $\alpha$ for the transmitted and reflected light.

\section{Perturbative Calculation of $g^{\left(2\right)}$\label{g2}}

\subsection{Correlation functions of the transmitted and reflected light.}

The second-order correlation function characterizes the bunched $\left(g^{\left(2\right)}\left(0\right)>\left(g^{\left(2\right)}\left(\tau\right)\right)\right)$
or anti-bunched $\left(g^{\left(2\right)}\left(0\right)<\left(g^{\left(2\right)}\left(\tau\right)\right)\right)$
nature of the light transmitted or reflected by the cavity. For the
transmitted light on the right side $(R)$ of the cavity, one has
by definition $g_{\mathrm{t}}^{\left(2\right)}\left(0\right)\equiv\left\langle a_{out}^{\left(R\right)\dagger}a_{out}^{\left(R\right)\dagger}a_{out}^{\left(R\right)}a_{out}^{\left(R\right)}\right\rangle /\left\langle a_{out}^{\left(R\right)\dagger}a_{out}^{\left(R\right)}\right\rangle ^{2}$,
where $a_{out}^{\left(R\right)}$ is the transmitted mode field annihilation
operator, and all averages should be evaluated in the steady state.
From the input-output relations \cite{Walls}, one gets $a_{out}^{\left(R\right)}+a_{in}^{\left(R\right)}=\sqrt{2\gamma_{c}^{\left(R\right)}}a$,
and hence 
\[
g_{\mathrm{t}}^{\left(2\right)}\left(0\right)=\left\langle a^{\dagger}a^{\dagger}aa\right\rangle /\left\langle a^{\dagger}a\right\rangle ^{2}.
\]
For the reflected light on the left side $(L)$ of the cavity, one
gets $g_{\mbox{r}}^{\left(2\right)}\left(0\right)\equiv\left\langle a_{out}^{\left(L\right)\dagger}a_{out}^{\left(L\right)\dagger}a_{out}^{\left(L\right)}a_{out}^{\left(L\right)}\right\rangle /\left\langle a_{out}^{\left(L\right)\dagger}a_{out}^{\left(L\right)}\right\rangle ^{2}$,
where $a_{out}^{\left(L\right)}$ is the reflected mode field annihilation
operator. Similarly, by using the input-output relation 
\[
a_{out}^{\left(L\right)}+a_{in}^{\left(L\right)}-\mathrm{i}\frac{\alpha}{\sqrt{2\gamma_{c}^{\left(L\right)}}}=\sqrt{2\gamma_{c}^{\left(L\right)}}a
\]
for the left mirror, one gets
\begin{eqnarray*}
 &  & \left\langle a_{out}^{\left(L\right)\dagger}a_{out}^{\left(L\right)\dagger}a_{out}^{\left(L\right)}a_{out}^{\left(L\right)}\right\rangle =\left(2\gamma_{c}^{\left(L\right)}\right)^{2}\left\langle a^{\dagger}a^{\dagger}aa\right\rangle +\\
 &  & \;\;\;\;4\mathrm{i}\alpha\gamma_{c}^{\left(L\right)}\left[\left\langle a^{\dagger}a^{\dagger}a\right\rangle -\left\langle a^{\dagger}aa\right\rangle \right]+\mathrm{i}\frac{\alpha^{3}}{\gamma_{c}^{\left(L\right)}}\left(\left\langle a^{\dagger}\right\rangle -\left\langle a\right\rangle \right)+\\
 &  & \;\;\;\;\alpha^{2}\left(4\left\langle a^{\dagger}a\right\rangle -\left\langle a^{\dagger}a^{\dagger}\right\rangle +\left\langle aa\right\rangle \right)+\frac{\alpha^{4}}{\left(2\gamma_{c}^{\left(L\right)}\right)^{2}}\\
 &  & \left\langle a_{out}^{\left(L\right)\dagger}a_{out}^{\left(L\right)}\right\rangle =2\gamma_{c}^{\left(L\right)}\left\langle a^{\dagger}a\right\rangle +\mathrm{i}\alpha\left(\left\langle a^{\dagger}\right\rangle -\left\langle a\right\rangle \right)+\frac{\alpha^{2}}{2\gamma_{c}^{\left(L\right)}}
\end{eqnarray*}

\subsection{Factorization in the perturbative limit}

In the whole paper, we shall restrict ourselves to the low excitation
regime, \emph{i.e.} to low values of the feeding parameter $\alpha$.
We therefore seek $g^{\left(2\right)}\left(0\right)$ at the lowest
non-vanishing order in $\alpha$: this requires to evaluate $\left\langle a^{\dagger}a^{\dagger}aa\right\rangle $,
$\left\langle a^{\dagger}a^{\dagger}a\right\rangle $ and $\left\langle a^{\dagger}a\right\rangle $
at the fourth, third and second orders, respectively. This task is
greatly simplified by the following remarkable factorization property,
established in Appendix \ref{Factorization}, 
\begin{eqnarray*}
 &  & \left\langle a^{\dagger}\left(t\right)a\left(t\right)\right\rangle ^{\left(2\right)}=\left\langle a^{\dagger}\left(t\right)\right\rangle ^{\left(1\right)}\left\langle a\left(t\right)\right\rangle ^{\left(1\right)}\\
 &  & \left\langle a^{\dagger}\left(t_{2}\right)a^{\dagger}\left(t_{1}\right)a\left(t_{1}\right)\right\rangle ^{\left(3\right)}=\left\langle a^{\dagger}\left(t_{2}\right)a^{\dagger}\left(t_{1}\right)\right\rangle ^{\left(2\right)}\times\left\langle a\left(t_{1}\right)\right\rangle ^{\left(1\right)}\\
 &  & \left\langle a^{\dagger}\left(t_{2}\right)a^{\dagger}\left(t_{1}\right)a\left(t_{1}\right)a\left(t_{2}\right)\right\rangle ^{\left(4\right)}=\\
 &  & \;\;\;\;\;\;\;\;\;\;\;\;\;\;\;\;\;\;\;\;\;\;\;\;\;\;\;\;\;\;\;\;\left\langle a^{\dagger}\left(t_{2}\right)a^{\dagger}\left(t_{1}\right)\right\rangle ^{\left(2\right)}\times\left\langle a\left(t_{1}\right)a\left(t_{2}\right)\right\rangle ^{\left(2\right)}
\end{eqnarray*}
\\
where the superscript $(k)$ denotes the order in $\alpha$ to which
quantities are calculated. Therefore, for instance, for the transmitted
light, 
\[
g_{\mathrm{t}}^{\left(2\right)}\left(0\right)=\left(\left\langle a^{\dagger}a^{\dagger}\right\rangle ^{\left(2\right)}\left\langle aa\right\rangle ^{\left(2\right)}\right)/\left(\left\langle a^{\dagger}\right\rangle ^{\left(1\right)}\left\langle a\right\rangle ^{\left(1\right)}\right)^{2}
\]
and we merely need to determine $\left\langle a\right\rangle ^{\left(1\right)}$
and $\left\langle a^{2}\right\rangle ^{\left(2\right)}$. Note that
the factorization does not apply to products of the kind $\left\langle a^{2}\right\rangle ^{\left(2\right)}$,
so that $\left\langle a^{2}\right\rangle ^{\left(2\right)}\neq\left\langle a\right\rangle ^{\left(1\right)}\left\langle a\right\rangle ^{\left(1\right)}$.

The mean values $\left\langle a\right\rangle ^{\left(1\right)}$and
$\left\langle \sigma_{ge}^{\left(i\right)}\right\rangle ^{\left(1\right)}$
are readily obtained through taking the steady state of the first-order
averaged Heisenberg equations Eqs. (\ref{HL1}-\ref{HL4}) 
\begin{align}
\left\langle a\right\rangle ^{\left(1\right)} & =\frac{\alpha}{D_{c}-\frac{g^{2}N}{\left(D_{e}-\frac{\Omega_{cf}^{2}}{4D_{r}}\right)}}\label{a1}\\
\left\langle \sigma_{ge}^{\left(i\right)}\right\rangle ^{\left(1\right)} & =\frac{\alpha g}{D_{c}\left(D_{e}-\frac{\Omega_{cf}^{2}}{4D_{r}}\right)-g^{2}N}\label{sigmage1}\\
\left\langle \sigma_{gr}^{\left(i\right)}\right\rangle ^{\left(1\right)} & =\frac{\alpha g\Omega_{cf}}{2D_{r}\left[D_{c}\left(D_{e}-\frac{\Omega_{cf}^{2}}{4D_{r}}\right)-g^{2}N\right]}\label{sigmagr1}
\end{align}

The second-order value $\left\langle a^{2}\right\rangle ^{\left(2\right)}$
is determined through solving the following closed system 
\begin{widetext}
\begin{eqnarray}
\left\langle a^{2}\right\rangle ^{\left(2\right)} & = & \frac{g\sqrt{N}}{D_{c}}\left\langle ab\right\rangle ^{\left(2\right)}+\frac{\alpha}{D_{c}}\left\langle a\right\rangle ^{\left(1\right)}\label{SO1}\\
\left\langle ab\right\rangle ^{\left(2\right)} & = & \frac{\Omega_{cf}}{2\left(D_{c}+D_{e}\right)}\left\langle ac\right\rangle ^{\left(2\right)}+\frac{g\sqrt{N}}{\left(D_{c}+D_{e}\right)}\left\langle aa\right\rangle ^{\left(2\right)}+\frac{g\sqrt{N}}{\left(D_{c}+D_{e}\right)}\left\langle bb\right\rangle ^{\left(2\right)}+\frac{\alpha}{\left(D_{c}+D_{e}\right)}\left\langle b\right\rangle ^{\left(1\right)}\label{SO2}\\
\left\langle ac\right\rangle ^{\left(2\right)} & = & \frac{g\sqrt{N}}{\left(D_{c}+D_{r}\right)}\left\langle bc\right\rangle ^{\left(2\right)}+\frac{\alpha}{\left(D_{c}+D_{r}\right)}\left\langle c\right\rangle ^{\left(1\right)}+\frac{\Omega_{cf}}{2\left(D_{c}+D_{r}\right)}\left\langle ab\right\rangle ^{\left(2\right)}\label{SO3}\\
\left\langle bb\right\rangle ^{\left(2\right)} & = & \frac{\Omega_{cf}}{2D_{e}}\left\langle bc\right\rangle ^{\left(2\right)}+\frac{g\sqrt{N}}{D_{e}}\left\langle ab\right\rangle ^{\left(2\right)}\label{SO4}\\
\left\langle bc\right\rangle ^{\left(2\right)} & = & \frac{\Omega_{cf}}{2\left(D_{e}+D_{r}\right)}\left\langle cc\right\rangle ^{\left(2\right)}+\frac{g\sqrt{N}}{\left(D_{e}+D_{r}\right)}\left\langle ac\right\rangle ^{\left(2\right)}+\frac{\Omega_{cf}}{2\left(D_{e}+D_{r}\right)}\left\langle bb\right\rangle ^{\left(2\right)}\label{SO5}\\
\left\langle cc\right\rangle ^{\left(2\right)} & = & \frac{\Omega_{cf}g\sqrt{N}}{2}K\left\langle ac\right\rangle ^{\left(2\right)}+\frac{\Omega_{cf}^{2}g\sqrt{N}}{4D_{e}}K\left\langle ab\right\rangle ^{\left(2\right)}\label{SO6}
\end{eqnarray}
deduced from Eqs. (\ref{HL1}-\ref{HL4}) under the assumption of
an homogeneous atomic medium, whose consequences are detailed in Appendix
\ref{a2}. In this system, we introduced the collective atomic operators
\[
b\equiv\frac{1}{\sqrt{N}}\sum_{i}\sigma_{ge}^{\left(i\right)}\;\;\;\;\; c\equiv\frac{1}{\sqrt{N}}\sum_{i}\sigma_{gr}^{\left(i\right)}.
\]

\end{widetext}

We note that the first-order mean values $\left\langle a\right\rangle ^{\left(1\right)}$,
$\left\langle b\right\rangle ^{\left(1\right)}$ and $\left\langle c\right\rangle ^{\left(1\right)}$
which appear in Eqs. (\ref{SO1}, \ref{SO2}, \ref{SO3}), respectively,
have been computed in Eqs. (\ref{a1}, \ref{sigmage1}, \ref{sigmagr1}).
The $K$ coefficient is approximately given by (see Appendix \ref{a2}
for details) 
\begin{equation}
K\approx\frac{1}{\left(D_{e}+D_{r}-\frac{\Omega_{cf}^{2}}{4D_{e}}\right)D_{r}-\frac{\Omega_{b}^{2}}{4}}\left(1-\frac{V_{b}}{V}\right)\label{K}
\end{equation}
where 
\begin{equation}
V_{b}=\frac{\sqrt{2}\pi^{2}}{3}\sqrt{\frac{-C_{6}}{D_{r}-\Omega_{cf}^{2}/\left(4(D_{e}+D_{r})-\frac{\Omega_{cf}^{2}}{D_{e}}\right)}}\label{Vb}
\end{equation}
will be interpreted as the Rydberg bubble volume in the dispersive
regime in the next section. Though it is too cumbersome to be reproduced
here, the solution for $\left\langle a^{2}\right\rangle ^{\left(2\right)}$
is simply obtained by matrix inversion, and the calculation of $g_{\mathrm{t}}^{\left(2\right)}\left(0\right)$
and $g_{\mbox{r}}^{\left(2\right)}\left(0\right)$ can be straightforwardly
programmed, e.g. in Mathematica.

As it has been the case for $g_{\mathrm{t,r}}^{\left(2\right)}\left(0\right)$,
the calculation of the time-dependent correlation function $g_{\mbox{t,r}}^{\left(2\right)}\left(\tau\right)\equiv\left\langle a^{\dagger}\left(t\right)a^{\dagger}\left(t+\tau\right)a\left(t+\tau\right)a\left(t\right)\right\rangle /\left\langle a^{\dagger}a\right\rangle ^{2}$
is greatly simplified by the factorization property derived in Appendix
\ref{Factorization}, since we simply need to determine the quantity
$\left\langle a\left(t+\tau\right)a\left(t\right)\right\rangle $.
From Eqs. (\ref{HL1}-\ref{HL4}), one easily deduces the following
differential system, at the lowest order in $\alpha$, 
\begin{eqnarray*}
 &  & \frac{d}{d\tau}\left(\begin{array}{c}
\left\langle a\left(t+\tau\right)a\left(t\right)\right\rangle \\
\left\langle b\left(t+\tau\right)a\left(t\right)\right\rangle \\
\left\langle c\left(t+\tau\right)a\left(t\right)\right\rangle 
\end{array}\right)=-\mbox{i}\alpha\left\langle a\right\rangle \left(\begin{array}{c}
1\\
0\\
0
\end{array}\right)+\\
 &  & -\mathrm{i}\left(\begin{array}{ccc}
-D_{c} & g\sqrt{N} & 0\\
g\sqrt{N} & -D_{e} & \frac{\Omega_{cf}}{2}\\
0 & \frac{\Omega_{cf}}{2} & -D_{r}
\end{array}\right)\left(\begin{array}{c}
\left\langle a\left(t+\tau\right)a\left(t\right)\right\rangle \\
\left\langle b\left(t+\tau\right)a\left(t\right)\right\rangle \\
\left\langle c\left(t+\tau\right)a\left(t\right)\right\rangle 
\end{array}\right)
\end{eqnarray*}
which, together with the initial condition 
\[
\left(\begin{array}{c}
\left\langle a\left(t+\tau\right)a\left(t\right)\right\rangle \\
\left\langle b\left(t+\tau\right)a\left(t\right)\right\rangle \\
\left\langle c\left(t+\tau\right)a\left(t\right)\right\rangle 
\end{array}\right)_{\tau\equiv0}=\left(\begin{array}{c}
\left\langle aa\right\rangle ^{\left(2\right)}\\
\left\langle ba\right\rangle ^{\left(2\right)}\\
\left\langle ca\right\rangle ^{\left(2\right)}
\end{array}\right)
\]
calculated above, allows us to determine $\left\langle a\left(t+\tau\right)a\left(t\right)\right\rangle $.
Again, though involved, the expressions are straightforward to obtain
and program.

\subsection{Application to an experimental case.}

\subsubsection{Dispersive regime.}

Let us now provide some numerical results obtained in the perturbative
approach described above. We first investigate the dispersive non-resonant
regime, addressed in our previous work \cite{GBB14}. To be specific,
we consider the same system, namely an ensemble of $^{87}$Rb atoms,
whose state space is restricted to the levels $\left|g\right\rangle =\left|5s_{\frac{1}{2}};F=2\right\rangle $,
$\left|e\right\rangle =\left|5p_{\frac{3}{2}};F=3\right\rangle $
and $\left|r\right\rangle =\left|95d_{\frac{5}{2}};F=4\right\rangle $.
The respective radiative decay rates are $\gamma_{e}=2\pi\times3$
MHz and $\gamma_{r}=2\pi\times0.03$ MHz, the cavity decay rate is
$\gamma_{c}=2\pi\times1$ MHz, the volume of the sample is $V=40\pi\times15^{2}\;\mu$m$^{3}$,
the sample density $n_{at}=0.4\;\mu$m$^{-3}$, and the cooperativity
$C=g^{2}N/(2\gamma_{e}\gamma_{c})=1000$.

The other parameters take the same values as in \cite{GBB14}: in
units of $\gamma_{e}$, the control laser Rabi frequency is $\Omega_{cf}=10$,
the detuning of the intermediate level is $\Delta_{e}=-35$, the detuning
of the Rydberg level is $\Delta_{r}=0.4$, the cavity feeding rate
is $\alpha=0.01$, and the Van der Waals coefficient is $C_{6}=-8.83\times10^{6}\gamma_{e}\;\mu$m$^{6}$.
For these parameters, the maximal average number of photons in the
cavity is obtained for the cavity detuning $\Delta_{c}^{\left(0\right)}=-6.15206\;\gamma_{e}$
which is taken as a reference.

Let us note however that in real experimental conditions, the atoms
undergo not only radiative damping, but are also subject to extra
dephasing $\gamma_{d}$ on the Rydberg-ground state transition, due
to laser frequency and intensity noise. This additional dephasing
cannot be modeled in the Hamiltonian formalism presented in Appendix
\ref{Hamiltonian}, and thus the demonstration given in Appendix \ref{Factorization}
for the factorization of mean values does not apply any more. However,
since the radiative coherence damping is $\gamma_{r}\approx0.01\,\gamma_{e}$,
the additional damping is $\gamma_{d}\approx0.15\,\gamma_{e}$, and
the total number of atoms in the sample is $N\approx10^{4}$, the
experimental parameters satisfy the condition $\gamma_{r}\ll\gamma_{d}\ll N\gamma_{r}$.
Under these circumstances, it is shown in Appendix \ref{FactDeph}
that the factorization remains valid, provided that the coherence
radiative damping $\gamma_{r}$ is replaced by the dephasing rate
$\gamma_{d}$ in the equations.

\begin{figure}
\begin{centering}
\includegraphics[width=8cm]{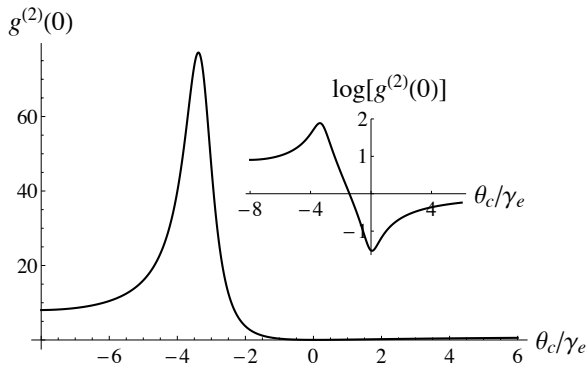} 
\par\end{centering}

\protect\caption{Second-order correlation function $g_{\mbox{t}}^{\left(2\right)}\left(0\right)$
for the transmitted light in the dispersive regime considered in \cite{GBB14}
as a function of the renormalized cavity detuning $\theta_{c}/\gamma_{e}\equiv\left(\Delta_{c}-\Delta_{c}^{\left(0\right)}\right)/\gamma_{e}$
where $\Delta_{c}^{\left(0\right)}$ is the detuning of the linear
cavity. The shape of the plot is in good qualitative agreement with
the results of the previous model. Inset : the same plot in logarithmic
scale (bunching and antibunching peaks are more clearly visible).}

\label{Figg2disp} 
\end{figure}

Under these conditions, Figure \ref{Figg2disp} shows the second-order
correlation function $g_{\mbox{t}}^{\left(2\right)}\left(0\right)$
as a function of the reduced cavity detuning $\theta_{c}\equiv(\Delta_{c}-\Delta_{c}^{\left(0\right)})/\gamma_{e}$,
to be compared with Fig. 2 a) in \cite{GBB14}. The two plots are
in good qualitative agreement, but the position of the bunching peak
is shifted from $\theta_{c}\approx-5$ to $\theta_{c}\approx-3.5$,
for the same parameters. This basically originates from the definition
of $V_{b}$ in \cite{GBB14}, differing from the present one by a
factor $\sqrt{2}$.

\subsubsection{Resonant case}

After checking that the present work confirms our previous results,
obtained in the dispersive regime, let us consider the resonant case,
which could not be treated before. As a new set of paramenters, we
take $\Delta_{c}=\Delta_{e}=\Delta_{r}=0$, and we assume that $\gamma_{c}^{\left(R\right)}\ll\gamma_{c}^{\left(L\right)}$.
We also choose a higher principal number $n=100$ for the Rydberg
level, for which $\gamma_{r}=0.1\gamma_{e}$. In addition, we fix
$\gamma_{c}=0.3\gamma_{e}$, $C=\frac{g^{2}N}{2\gamma_{e}\gamma_{c}}\approx30$
and $V=50\pi\times20\times20\mu\mbox{m}^{3}$. In this regime, $V_{b}\approx\frac{\sqrt{2}\pi^{2}}{3}\sqrt{\frac{-C_{6}}{D_{e}}}$
is enhanced, therefore magnified non-linear effects are expected.

\begin{figure}[h]
\begin{centering}
\includegraphics[width=8cm]{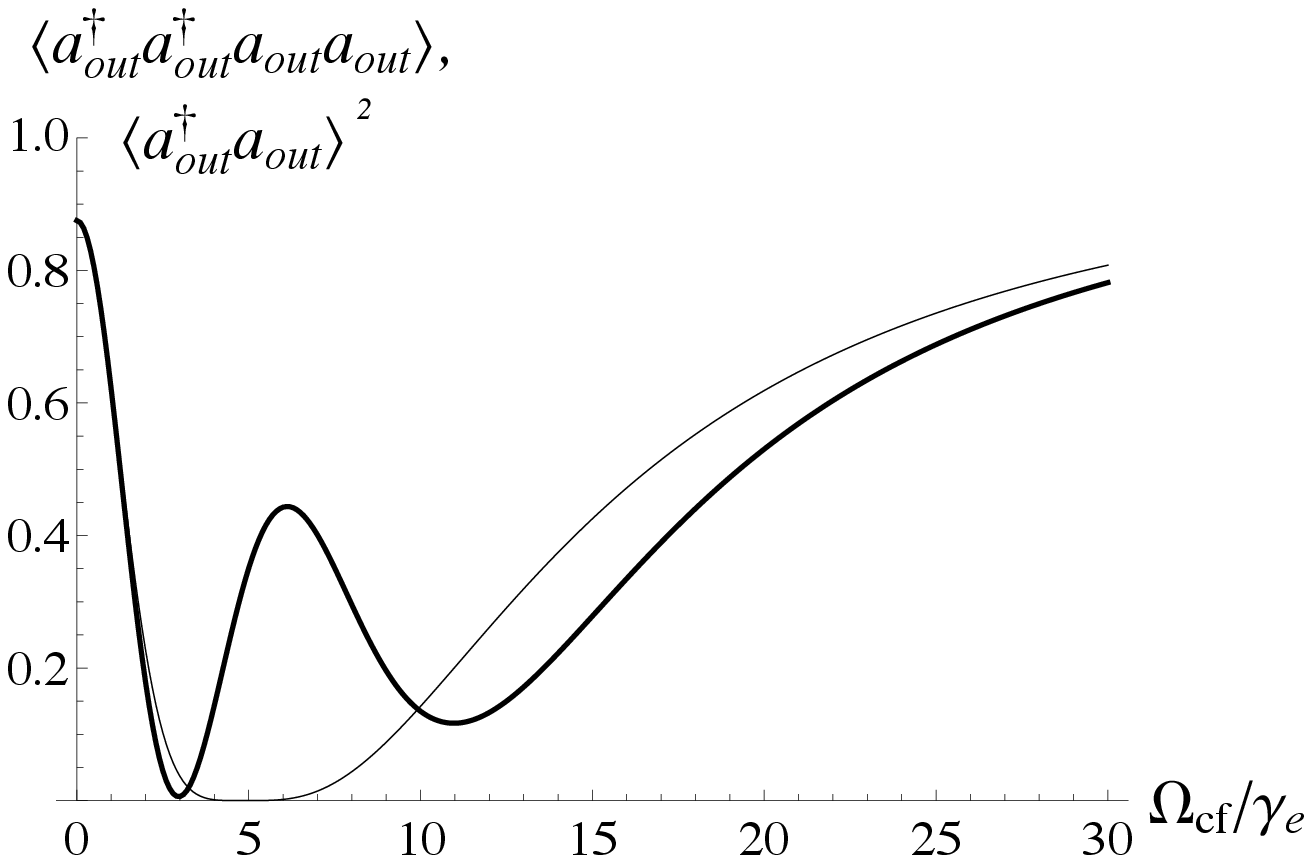} 
\par\end{centering}

\protect\caption{\emph{Resonant case }$\Delta_{c}=\Delta_{e}=\Delta_{r}=0$. The quantities
$\left\langle a_{out}^{\left(L\right)\dagger}a_{out}^{\left(L\right)\dagger}a_{out}^{\left(L\right)}a_{out}^{\left(L\right)}\right\rangle $
(thick line) and $\left\langle a_{out}^{\left(L\right)\dagger}a_{out}^{\left(L\right)}\right\rangle $
(thin line), renormalized by the intensity of the incoming light,
are represented as functions of the normalized control field Rabi
frequency $\Omega_{cf}/\gamma_{e}$. For $\Omega_{cf}=2\sqrt{\gamma_{e}\gamma_{r}\left(2C-1\right)}\approx5\gamma_{e}$,
photon pairs are reflected, \emph{i.e.} $\left\langle a_{out}^{\left(L\right)\dagger}a_{out}^{\left(L\right)\dagger}a_{out}^{\left(L\right)}a_{out}^{\left(L\right)}\right\rangle \protect\neq0$,
while single photons are absorbed, \emph{i.e.} $\left\langle a_{out}^{\left(L\right)\dagger}a_{out}^{\left(L\right)}\right\rangle \approx0$.}

\label{FigRes} 
\end{figure}

\begin{figure}[H]
\begin{centering}
\includegraphics[width=8cm]{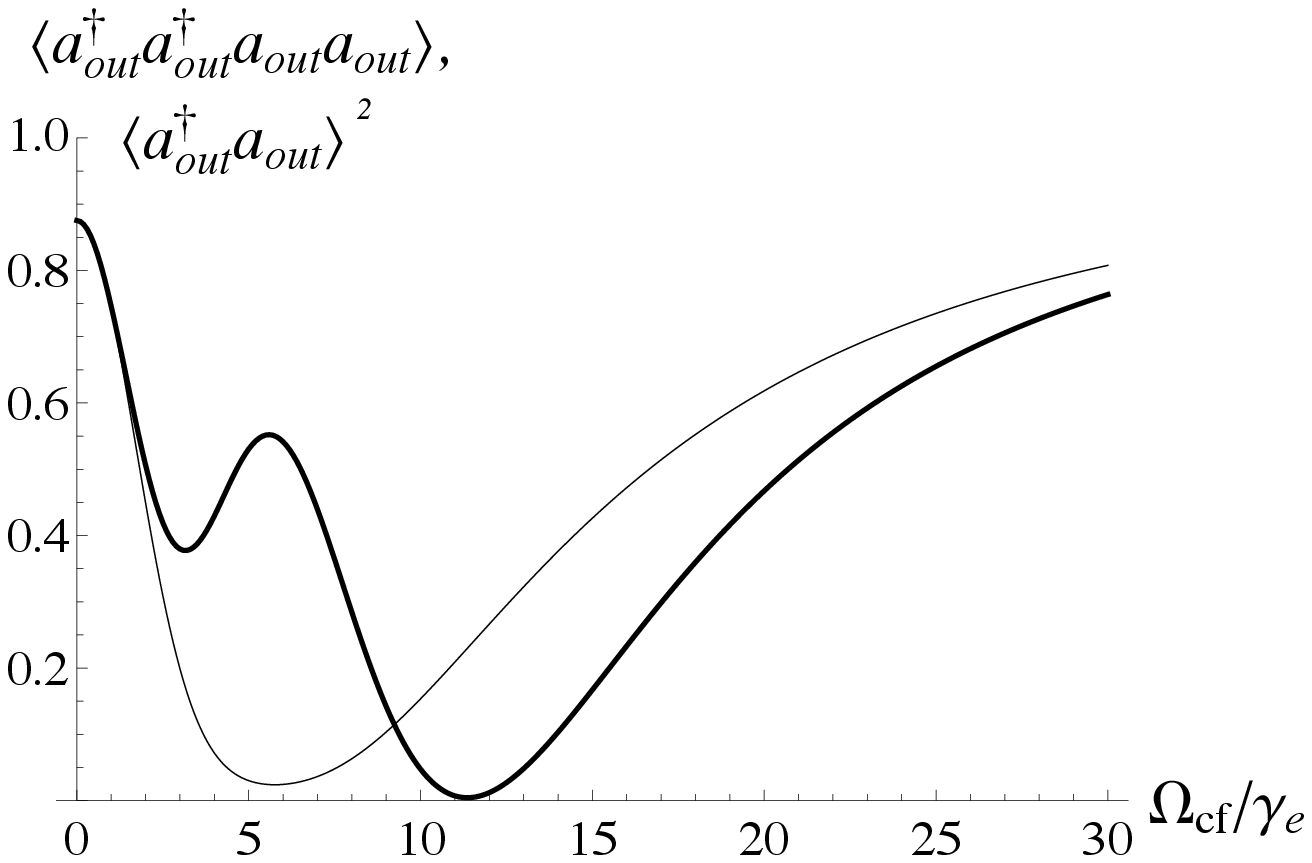} 
\par\end{centering}

\protect\caption{\emph{Slightly detuned case }$\Delta_{c}=0,$ $\Delta_{e}=-2\gamma_{e}$,
$\Delta_{r}=-0.1\gamma_{e}$. The quantities $\left\langle a_{out}^{\left(L\right)\dagger}a_{out}^{\left(L\right)\dagger}a_{out}^{\left(L\right)}a_{out}^{\left(L\right)}\right\rangle $
(thick line) and $\left\langle a_{out}^{\left(L\right)\dagger}a_{out}^{\left(L\right)}\right\rangle $
(thin line), renormalized by the intensity of the incoming light,
are represented as functions of the normalized control field Rabi
frequency $\Omega_{cf}/\gamma_{e}$. For $\Omega_{cf}\approx11\gamma_{e}$,
photon pairs are absorbed, \emph{i.e.} $\left\langle a_{out}^{\left(L\right)\dagger}a_{out}^{\left(L\right)\dagger}a_{out}^{\left(L\right)}a_{out}^{\left(L\right)}\right\rangle =0$,
while single photons are reflected, \emph{i.e.} $\left\langle a_{out}^{\left(L\right)\dagger}a_{out}^{\left(L\right)}\right\rangle \protect\neq0$.
\label{FigNearRes}}
\end{figure}

As can be seen on Figure \ref{FigRes}, there exists a value for which
single photons are mostly absorbed $\left\langle a_{out}^{\left(L\right)\dagger}a_{out}^{\left(L\right)}\right\rangle =0$,
while pairs are reflected $\left\langle a_{out}^{\left(L\right)\dagger}a_{out}^{\left(L\right)\dagger}a_{out}^{\left(L\right)}a_{out}^{\left(L\right)}\right\rangle \neq0$:
this value can be computed and is found to be 
\[
\Omega_{cf}=2\sqrt{\gamma_{e}\gamma_{r}\left(2C-1\right)}=2\gamma_{e}\sqrt{6}\approx5\gamma_{e}
\]
On the contrary, in a slightly detuned case, \emph{i.e.} for $\Delta_{e}=-2\gamma_{e}$
and $\Delta_{r}=-0.1\gamma_{e}$, the other parameters remaining the
same, one observes that around $\Omega_{cf}\approx11\gamma_{e}$ pairs
are absorbed $\left\langle a_{out}^{\left(L\right)\dagger}a_{out}^{\left(L\right)\dagger}a_{out}^{\left(L\right)}a_{out}^{\left(L\right)}\right\rangle =0$
while single photons are reflected $\left\langle a_{out}^{\left(L\right)\dagger}a_{out}^{\left(L\right)}\right\rangle \neq0$
(see Fig. \ref{FigNearRes}).

These new features are specific of the near-resonant regime, and were
not present in our previous work. They may be interpreted as different
impedance matching conditions for single photons and for pairs, leading
to very large non-linear losses, acting at the single photon level.

To conclude this section, we described how to obtain the exact and
analytic expression of the correlation function in the low excitation
regime, valid not only in the dispersive regime but even in the resonant
case. Though exact and computable, the expressions we get are too
cumbersome to be displayed here and do not easily lend themselves
to physical interpretation. In the next section, we introduce an effective
non-linear three-boson model which allows us to derive the same results
to the lowest order, and has also the advantage of being physically
more transparent.

\section{Effective non-linear three-boson model\label{ModelHam}}

\subsection{Non-linear absorption and dispersion in the quantum regime.}

We consider a system of three bosons of respective annihilation operators
$a$, $b$ and $c$, whose non-linear Hamiltonian is given by

\begin{align*}
H & =-\Delta_{c}a^{\dagger}a+\alpha\left(a+a^{\dagger}\right)-\Delta_{e}b^{\dagger}b\\
 & -\Delta_{r}c^{\dagger}c+g\sqrt{N}\left(ab^{\dagger}+b^{\dagger}a\right)\\
 & +\frac{\Omega_{cf}}{2}\left(bc^{\dagger}+b^{\dagger}c\right)+\frac{\kappa_{r}}{2}c^{\dagger}c^{\dagger}cc
\end{align*}
We moreover assume that the $c$-boson is coupled to a \emph{non-linear}
bath whose action on the system is represented by the following non-linear
dissipation operator, acting on the density matrix $\rho$ of the
system 
\[
\mathcal{D}\left[\rho\right]=\frac{\kappa_{i}}{2}\left\{ 2cc\rho c^{\dagger}c^{\dagger}-c^{\dagger}c^{\dagger}cc\rho-\rho c^{\dagger}c^{\dagger}cc\right\} 
\]
Here, all parameters, in particular $\kappa_{r}$ and $\kappa_{i}$,
are assumed real. From the full Liouville-von Neumann equation of
the system $\partial_{t}\rho=-\frac{\mathrm{i}}{\hbar}\left[H,\rho\right]+\mathcal{D}\left[\rho\right]$
one readily derives the following Bloch equations

\begin{eqnarray*}
\frac{d}{dt}\left\langle a\right\rangle  & = & \mathrm{i}D_{c}\left\langle a\right\rangle -\mathrm{i}\alpha-\mathrm{i}g\sqrt{N}\left\langle b\right\rangle \\
\frac{d}{dt}\left\langle b\right\rangle  & = & \mathrm{i}D_{e}\left\langle b\right\rangle -\mathrm{i}g\sqrt{N}\left\langle a\right\rangle -\mathrm{i}\frac{\Omega_{cf}}{2}\left\langle c\right\rangle \\
\frac{d}{dt}\left\langle c\right\rangle  & = & \mathrm{i}D_{r}\left\langle c\right\rangle -\mathrm{i}\frac{\Omega_{cf}}{2}\left\langle b\right\rangle -\mathrm{i}\kappa\left\langle c^{+}cc\right\rangle 
\end{eqnarray*}
where we introduced the notation $\kappa\equiv\kappa_{r}-\mathrm{i}\kappa_{i}$.
From this set of equations, one gets the same steady state value $\left\langle a\right\rangle ^{\left(1\right)}$
as in Eq. (\ref{a1}). At the second order in $\alpha$, the set of
equations for two-operator steady-state averages is derived in the
same way (here we omit superscripts $^{\left(1,2\right)}$ for simplicity) 
\begin{widetext}
\begin{eqnarray*}
\left\langle aa\right\rangle  & = & \frac{g\sqrt{N}}{D_{c}}\left\langle ab\right\rangle +\frac{\alpha}{D_{c}}\left\langle a\right\rangle \\
\left\langle ab\right\rangle  & = & \frac{\Omega_{cf}}{2\left(D_{c}+D_{e}\right)}\left\langle ac\right\rangle +\frac{g\sqrt{N}}{\left(D_{c}+D_{e}\right)}\left\langle aa\right\rangle +\frac{g\sqrt{N}}{\left(D_{c}+D_{e}\right)}\left\langle bb\right\rangle +\frac{\alpha}{\left(D_{c}+D_{e}\right)}\left\langle b\right\rangle \\
\left\langle ac\right\rangle  & = & \frac{g\sqrt{N}}{\left(D_{c}+D_{r}\right)}\left\langle bc\right\rangle +\frac{\alpha}{\left(D_{c}+D_{r}\right)}\left\langle c\right\rangle +\frac{\Omega_{cf}}{2\left(D_{c}+D_{r}\right)}\left\langle ab\right\rangle \\
\left\langle bb\right\rangle  & = & \frac{\Omega_{cf}}{2D_{e}}\left\langle bc\right\rangle +\frac{g\sqrt{N}}{D_{e}}\left\langle ab\right\rangle \\
\left\langle bc\right\rangle  & = & \frac{\Omega_{cf}}{2\left(D_{e}+D_{r}\right)}\left\langle cc\right\rangle +\frac{g\sqrt{N}}{\left(D_{e}+D_{r}\right)}\left\langle ac\right\rangle +\frac{\Omega_{cf}}{2\left(D_{e}+D_{r}\right)}\left\langle bb\right\rangle \\
\left\langle cc\right\rangle  & = & \frac{\Omega_{cf}}{2\left(D_{r}-\frac{\kappa}{2}\right)}\left\langle bc\right\rangle 
\end{eqnarray*}

\end{widetext}

which agrees with Eqs. (\ref{SO1}-\ref{SO6}) but for the last equation.
If, however, we eliminate$\left\langle bc\right\rangle $ and $\left\langle bb\right\rangle $
from the last three equations, one obtains

\begin{eqnarray*}
\left\langle cc\right\rangle  & = & \frac{1}{\left(D_{r}-\frac{\kappa}{2}\right)\left(D_{r}+D_{e}-\frac{\Omega_{cf}^{2}}{4D_{e}}\right)-\frac{\Omega_{cf}^{2}}{4}}\\
 &  & \times\frac{\Omega_{cf}g\sqrt{N}}{2}\left(\left\langle ac\right\rangle +\frac{\Omega_{cf}}{2D_{e}}\left\langle ab\right\rangle \right)
\end{eqnarray*}
which can be identified with Eq. (\ref{SO6}) provided that 
\begin{align*}
K & =\frac{1}{\left(D_{r}-\frac{\kappa}{2}\right)\left(D_{r}+D_{e}-\frac{\Omega_{cf}^{2}}{4D_{e}}\right)-\frac{\Omega_{cf}^{2}}{4}}
\end{align*}
which, upon recalling Eq. (\ref{K}), yields

\begin{align*}
\kappa & =2\left(\frac{V_{b}}{V-V_{b}}\right)\left(\frac{\Omega_{cf}^{2}}{4\left(D_{r}+D_{e}-\frac{\Omega_{cf}^{2}}{4D_{e}}\right)}-D_{r}\right)
\end{align*}
We obtain thus the analytic expressions of the parameters $\kappa_{r}$
and $\kappa_{i}$, respectively characterizing the non-linear dispersion
and absorption of the $c$-boson, which make our model system precisely
reproduce the results of the original problem in the steady state
and in the lowest order of the feeding parameter $\alpha$.

\subsection{Discussion.}

Let us now investigate the physical content of the previous model
by considering two limiting cases.

In the dispersive regime addressed in our previous work \cite{GBB14},
$\left|D_{e,r}\right|\gg\Omega_{cf}$, whence $V_{b}\approx\frac{\sqrt{2}\pi^{2}}{3}\sqrt{\frac{\left|C_{6}\right|}{\Delta_{r}}}$,
$\kappa_{r}\approx-\frac{2\Delta_{r}}{\left(N_{b}-1\right)}$ and
$\kappa_{i}\approx0$, where we introduced $N_{b}\equiv\frac{V}{V_{b}}$.
This result agrees with what we previously obtained in the Rydberg
bubble approximation \cite{GBB14} and therefore confirms its validity:
we observe a shift due to the non-linear dispersive behavior of the
$c$-boson, but no non-linear absorption since the intermediate level
is too far detuned. Moreover, in the bubble picture, the parameter
$N_{b}$ was interpreted as the number of Rydberg bubbles the sample
may accommodate; as suggested above, this allows to interpret $V_{b}$
as the bubble volume.

If we now go to the opposite regime, \emph{i.e.} the resonant case
for which $\Delta_{e}=\Delta_{r}=0$, $\gamma_{e}\gg\gamma_{r}$ and
$\Omega_{cf}^{2}\gg\gamma_{e}^{2}$, we obtain $V_{b}\approx\frac{\pi^{2}}{3}\left(1-\mathrm{i}\right)\sqrt{\frac{\left|C_{6}\right|}{\gamma_{e}}}$
and therefore the non-linearity parameters are 
\[
\kappa_{r}=-\kappa_{i}\approx-\frac{2\pi^{2}}{3V}\sqrt{\gamma_{e}\left|C_{6}\right|}
\]
We now have both dispersion \emph{and} absorption. From the expression
of $\kappa_{i}$, it is clear that absorption results from an interplay
of the spontaneous emission from the intermediate state and the Rydberg-Rydberg
interactions.

\section{Conclusion \label{sec:Conclusion}}

In this article, we have studied the strong quantum optical non-linearities
induced by Rydberg-Rydberg van der Waals interactions in an atomic
medium. We provided a new perturbative treatment of the problem, based
on the factorization of the lowest perturbative order of operator
product averages. Though being 
purely radiative damping, this factorization property is approximately
preserved in the presence of 
to, e.g., laser frequency and intensity noise, as it is the case in
our experimental setup. Our perturbative calculations enabled us to
recover and extend our previous results: we could validate the approach
based on the Rydberg bubble picture, as well as investigate the resonant,
absorptive, regime. In particular, our numerical simulations showed
that strong Rydberg-induced non-linearities led to different impedance
matching conditions for single photons and photon pairs.

Moreover we proposed an effective model which leads to the same results
as the full calculation at the lowest order in the feeding parameter;
this model also sheds some light on the origin of the dispersion and
absorption, as well as makes a bridge between the Rydberg bubble and
perturbative approaches. In the future, we shall first try and take
advantage of our understanding of the system to investigate regimes
of parameters for which a photonic gate can be implemented. On the
other hand, we also plan to apply other methods, inspired from many-body
physics to the problem, in order to recover and further extend our
results. 
\begin{acknowledgments}
This work is supported by the European Union grants DELPHI (ERC \#246669
) and SIQS (FET \#600645). 
\end{acknowledgments}

\appendix

\section{The full Hamiltonian in the Rotating Wave Approximation\label{Hamiltonian}}

The full Hamiltonian of the system can be written under the form 
\begin{eqnarray*}
 &  & H=H_{at}+H_{cav}+H_{bath}+V_{at-cav}+V_{cav-bath}+V_{at-bath}\\
 &  & H_{at}\equiv\hbar\omega_{e}\sum_{n=1}^{N}\sigma_{ee}^{\left(n\right)}+\hbar\Omega_{cf}\cos\left(\omega_{cf}t\right)\sum_{n=1}^{N}\left(\sigma_{re}^{\left(n\right)}+\sigma_{er}^{\left(n\right)}\right)+\\
 &  & \;\;\;\;\;\;\;\;\;\;\;\;+\hbar\omega_{r}\sum_{n=1}^{N}\sigma_{rr}^{\left(n\right)}+\sum_{m<n=1}^{N}\hbar\kappa_{mn}\sigma_{rr}^{\left(m\right)}\sigma_{rr}^{\left(n\right)}\\
 &  & H_{cav}\equiv\hbar\omega_{c}a^{\dagger}a\\
 &  & H_{bath}\equiv\int d\omega\;\hbar\omega\left(b^{\dagger}b+c^{\dagger}c+d^{\dagger}d\right)\\
 &  & V_{at-cav}\equiv\sum_{n=1}^{N}\hbar g\left(a+a^{\dagger}\right)\left(\sigma_{eg}^{\left(n\right)}+\sigma_{ge}^{\left(n\right)}\right)\\
 &  & V_{cav-bath}\equiv\int d\omega\;\hbar g_{b}\left(b+b^{\dagger}\right)\left(a+a^{\dagger}\right)\\
 &  & V_{at-bath}\equiv\int d\omega\;\hbar g_{c}\left(c+c^{\dagger}\right)\left(\sigma_{eg}^{\left(n\right)}+\sigma_{ge}^{\left(n\right)}\right)+\\
 &  & \;\;\;\;\;\;\;\;\;\;\;\;\;\;\;\;\;\;\;\;\;\;\;\;\int d\omega\;\hbar g_{d}\left(d+d^{\dagger}\right)\left(\sigma_{rg}^{\left(n\right)}+\sigma_{rg}^{\left(n\right)}\right)
\end{eqnarray*}
where $\sigma_{\alpha\beta}\equiv\left|\alpha\right\rangle \left\langle \beta\right|$,
$\hbar\omega_{\alpha}$ is the energy of the atomic level $\left|\alpha\right\rangle $
for $\alpha=e,r$ (with the convention $\omega_{g}=0$), and $\kappa_{mn}\equiv C_{6}/\left\Vert \vec{r}_{m}-\vec{r}_{n}\right\Vert ^{6}$
denotes the van der Waals interaction between atoms in the Rydberg
level -- when atoms are in the ground or intermediate states, their
interactions are neglected. The operators $b\left(\omega\right)$,
$c\left(\omega\right)$ and $d\left(\omega\right)$ denoted simply
as $b$, $c$, $d$, are bath operators coupled to the cavity and
atomic operators with the respective coupling strengths $g_{b}\left(\omega\right)$,
$g_{c}\left(\omega\right)$ and $g_{d}\left(\omega\right)$.

We switch to the rotating frame defined by $\left|\psi\right\rangle \rightarrow\left|\tilde{\psi}\right\rangle =\exp\left(-\frac{\mathrm{i}t}{\hbar}H_{0}\right)$
where 
\begin{align*}
H_{0} & \equiv\hbar\omega_{p}a^{\dagger}a+\sum_{n=1}^{N}\left(\hbar\omega_{p}\sigma_{ee}^{\left(n\right)}+\hbar\left(\omega_{p}+\omega_{cf}\right)\sigma_{rr}^{\left(n\right)}\right)+\\
 & \int d\omega\left(\hbar\omega_{p}b^{\dagger}b+\hbar\left(\omega_{p}+\omega_{cf}\right)c^{\dagger}c+\hbar\omega_{p}d^{\dagger}d\right)
\end{align*}
and perform the Rotating Wave Approximation to get the new Hamiltonian
\begin{eqnarray*}
 &  & \tilde{H}=\tilde{H}_{at}+\tilde{H}_{cav}+\tilde{H}_{bath}+\tilde{V}_{at-cav}+\tilde{V}_{at-bath}+\tilde{V}_{cav-bath}\\
 &  & \tilde{H}_{at}=-\hbar\Delta_{e}\sum_{n=1}^{N}\sigma_{ee}^{\left(n\right)}+\sum_{m<n=1}^{N}\hbar\kappa_{mn}\sigma_{rr}^{\left(m\right)}\sigma_{rr}^{\left(n\right)}+\\
 &  & \;\;\;\;\;-\hbar\Delta_{r}\sum_{n=1}^{N}\sigma_{rr}^{\left(n\right)}+\frac{\hbar\Omega_{cf}}{2}\sum_{n=1}^{N}\left(\sigma_{re}^{\left(n\right)}+\sigma_{er}^{\left(n\right)}\right)\\
 &  & \tilde{H}_{cav}=-\hbar\Delta_{c}a^{\dagger}a\\
 &  & \tilde{H}_{bath}\approx\int d\omega\;\hbar\omega\left(b\left(\omega+\omega_{p}\right)\right)^{\dagger}b\left(\omega+\omega_{p}\right)+\\
 &  & \;\;\int d\omega\;\hbar\omega\sum_{n=1}^{N}\left(c_{n}\left(\omega+\omega_{p}+\omega_{cf}\right)\right)^{\dagger}c_{n}\left(\omega+\omega_{p}+\omega_{cf}\right)+\\
 &  & \;\;\;\;\ \int d\omega\;\hbar\omega\sum_{n=1}^{N}\left(d_{n}\left(\omega+\omega_{p}\right)\right)^{\dagger}d_{n}\left(\omega+\omega_{p}\right)\\
 &  & \tilde{V}_{a-c}\approx\sum_{n=1}^{N}\hbar g\left(a\sigma_{eg}^{\left(n\right)}+a^{\dagger}\sigma_{ge}^{\left(n\right)}\right)\\
 &  & \tilde{V}_{cav-bath}\approx\int d\omega\;\hbar g_{b}\left(\omega\right)\left[b\left(\omega\right)a^{\dagger}+\left(b\left(\omega\right)\right)^{\dagger}a\right]\\
 &  & \tilde{V}_{at-bath}\approx\sum_{n=1}^{N}\int d\omega\;\hbar g_{c}\left(\omega\right)\left[c_{n}\left(\omega\right)\sigma_{eg}^{\left(n\right)}+\left(c_{n}\left(\omega\right)\right)^{\dagger}\sigma_{ge}^{\left(n\right)}\right]+\\
 &  & \;\;\;\;\;\sum_{n=1}^{N}\int d\omega\;\hbar g_{d}\left(\omega\right)\left[d_{n}\left(\omega\right)\sigma_{rg}^{\left(n\right)}+\left(d_{n}\left(\omega\right)\right)^{\dagger}\sigma_{rg}^{\left(n\right)}\right]
\end{eqnarray*}
with the detunings $\Delta_{c}\equiv\left(\omega_{p}-\omega_{c}\right)$,
$\Delta_{e}\equiv\left(\omega_{p}-\omega_{e}\right)$, and $\Delta_{r}\equiv\left(\omega_{p}+\omega_{cf}-\omega_{r}\right)$.
It is important to note that the evolution under the Hamiltonian $\tilde{H}$
conserves the number of excitations.

\section{Factorization of correlation functions. \label{Factorization}}

We suppose that the bath interacting with the cavity is initially
in the following continuous-mode coherent state (incoming quasi-classical
field) 
\[
\left|\alpha\right\rangle =e^{-\frac{1}{2}\left\langle n\right\rangle }e^{\sqrt{\left\langle n\right\rangle }b_{\alpha}^{\dagger}}\left|0\right\rangle 
\]
where $\int\left|\alpha\left(t\right)\right|^{2}dt=\left\langle n\right\rangle $
and $b_{\alpha}^{\dagger}=\frac{1}{\sqrt{\left\langle n\right\rangle }}\int d\omega\alpha\left(\omega\right)b^{\dagger}\left(\omega\right)$
is a superposition of bath mode creation operators $b^{\dagger}\left(\omega\right)$
\cite{L00}. Note that with this definition, $b_{\alpha}$ is a bosonic
operator, \textbf{\emph{i.e.}} $\left[b_{\alpha},b_{\alpha}^{\dagger}\right]=1$.
The atoms and cavity field are initially in their ground state denoted
by $\left|G\right\rangle \equiv\left|g\ldots g\right\rangle \otimes\left|0\right\rangle $.

Let us consider, for instance, the quantity $\left\langle \alpha,G|a^{\dagger}\left(t_{1}\right)a^{\dagger}\left(t_{2}\right)a\left(t_{2}\right)a\left(t_{1}\right)|G,\alpha\right\rangle $,
for $t_{2}>t_{1}$, where $\left|G,\alpha\right\rangle $ denotes
the initial state of the whole system $\left\{ \mbox{atoms+cavity+baths}\right\} $,
the baths coupled to the atoms are supposed empty and their state
is not explicitly written, 
\begin{eqnarray}
 &  & \left\langle \alpha,G|a^{\dagger}\left(t_{1}\right)a^{\dagger}\left(t_{2}\right)a\left(t_{2}\right)a\left(t_{1}\right)|G,\alpha\right\rangle \label{producta}\\
 & = & e^{-\left\langle n\right\rangle }\sum_{k,l}\frac{\left\langle n\right\rangle ^{\frac{k+l}{2}}}{\sqrt{k!l!}}\left\langle k,G|a^{\dagger}\left(t_{1}\right)a^{\dagger}\left(t_{2}\right)a\left(t_{2}\right)a\left(t_{1}\right)|G,l\right\rangle \nonumber 
\end{eqnarray}
Expanding this expression with respect to $\left|\alpha\right|$ (which
is equivalent to expanding in the number of excitations present in
the system), one finds that the lowest non-vanishing contribution
is the fourth order term $k=l=2$. For the system considered the identity
operator can be represented in the following way $\mathcal{I}=\underset{i}{\bigotimes}\mathcal{I}_{i}$
where $\mathcal{I}_{i}=\sum_{q}\left|q_{i}\right\rangle \left\langle q_{i}\right|$
are the identity operators on each degree of freedom of the system,
and $\left|q_{i}\right\rangle $'s denote $q$ -th basis vector of
$i$ -th degree of freedom. Inserting this identity operator between
$a^{+}\left(t_{2}\right)$ and $a\left(t_{2}\right)$ of the quantity
(\ref{producta}) yields:

\begin{eqnarray}
e^{-\left\langle n\right\rangle }\sum_{k,l}\frac{\left\langle n\right\rangle ^{\frac{k+l}{2}}}{\sqrt{k!l!}}\left\langle k,G\right|a^{\dagger}\left(t_{1}\right)a^{\dagger}\left(t_{2}\right)\label{mean}\\
\left\{ \underset{i}{\bigotimes}\sum_{q}\left|q_{i}\right\rangle \left\langle q_{i}\right|\right\} a\left(t_{2}\right)a\left(t_{1}\right)\left|G,l\right\rangle \nonumber 
\end{eqnarray}
For the lowest non-vanishing term $k=2,l=2$: 
\begin{eqnarray*}
a\left(t_{2}\right)a\left(t_{1}\right)\left|G,2\right\rangle  & = & e^{\mathrm{i}\frac{\tilde{H}t_{2}}{\hbar}}ae^{\mathrm{i}\frac{\tilde{H}}{\hbar}\left(t_{1}-t_{2}\right)}a\left|G,2\left(t_{1}\right)\right\rangle 
\end{eqnarray*}
where $\left|G,2\left(t_{1}\right)\right\rangle \equiv e^{-\mathrm{i}\frac{\tilde{H}t_{1}}{\hbar}}\left|G,2\right\rangle $
(note that this state can contain excited atoms and/or cavity photons)\textbf{.}
The state $a\left|G,2\left(t_{1}\right)\right\rangle $ can at most
contain one excitation, and so can the state $e^{iH^{\prime}\left(t_{1}-t_{2}\right)}a\left|G,2\left(t_{1}\right)\right\rangle $
due to the conservation of excitation number. Hence $e^{iH^{\prime}t_{2}}ae^{iH^{\prime}\left(t_{1}-t_{2}\right)}a\left|G,2\left(t_{1}\right)\right\rangle $
can only have component on $\left|G,0\right\rangle $. Finally the
fourth order expression of (\ref{mean}) reads:

\begin{eqnarray*}
 &  & e^{-\left\langle n\right\rangle }\frac{\left\langle n\right\rangle ^{2}}{2}\left\langle 2,G|a^{\dagger}\left(t_{1}\right)a^{\dagger}\left(t_{2}\right)a\left(t_{2}\right)a\left(t_{1}\right)|G,2\right\rangle \\
 & = & e^{-\left\langle n\right\rangle }\frac{\left\langle n\right\rangle ^{2}}{2}\left\vert \left\langle 2,G|a^{\dagger}\left(t_{1}\right)a^{\dagger}\left(t_{2}\right)|G,0\right\rangle \right\vert ^{2}\\
 & = & \left\langle \alpha,G|a^{\dagger}\left(t_{1}\right)a^{\dagger}\left(t_{2}\right)|G,0\right\rangle _{2}\left\langle G,0|a\left(t_{2}\right)a\left(t_{1}\right)|G,\alpha\right\rangle _{2}
\end{eqnarray*}
where we used that $e^{-\frac{\left\langle n\right\rangle }{2}}\frac{\left\langle n\right\rangle }{\sqrt{2}}\left\langle 2,G|a^{\dagger}\left(t_{1}\right)a^{\dagger}\left(t_{2}\right)|G,0\right\rangle $
and $e^{-\frac{\left\langle n\right\rangle }{2}}\frac{\left\langle n\right\rangle }{\sqrt{2}}\left\langle 0,G|a\left(t_{2}\right)a\left(t_{1}\right)|G,2\right\rangle $
are equal to the second order expansion in $\left|\alpha\right|$
of quantities $\left\langle \alpha,G|a^{\dagger}\left(t_{1}\right)a^{\dagger}\left(t_{2}\right)|G,\alpha\right\rangle $
and $\left\langle \alpha,G|a^{\dagger}\left(t_{1}\right)a^{\dagger}\left(t_{2}\right)|G,\alpha\right\rangle $
respectively, which we denoted by $\left\langle ...\right\rangle _{2}$.

Thus to compute $\left\langle \alpha,G|a^{\dagger}\left(t_{1}\right)a^{\dagger}\left(t_{2}\right)a\left(t_{2}\right)a\left(t_{1}\right)|G,\alpha\right\rangle $
in the lowest order it is enough to calculate $\left\langle a\left(t_{2}\right)a\left(t_{1}\right)\right\rangle \equiv\left\langle \alpha,G|a\left(t_{2}\right)a\left(t_{1}\right)|G,\alpha\right\rangle $.
\\
 \\
 The same argument holds for more general mean values such as 
\begin{align*}
\left\langle \alpha,G\right|a^{\dagger}\left(t_{1}\right)a^{\dagger}\left(t_{2}\right)\dots a^{\dagger}\left(t_{p}\right)a\left(t_{p+1}\right)\dots\\
\dots a\left(t_{p+q-1}\right)a\left(t_{p+q}\right)\left|G,\alpha\right\rangle ^{\left(p+q\right)}
\end{align*}
and in particular 
\begin{align*}
\left\langle a^{\dagger}\left(t\right)a\left(t\right)\right\rangle ^{\left(2\right)} & =\left\langle a^{\dagger}\left(t\right)\right\rangle ^{\left(1\right)}\left\langle a\left(t\right)\right\rangle ^{\left(1\right)}\\
\left\langle a^{\dagger}\left(t_{2}\right)a^{\dagger}\left(t_{1}\right)a\left(t_{1}\right)\right\rangle ^{\left(3\right)} & =\left\langle a^{\dagger}\left(t_{2}\right)a^{\dagger}\left(t_{1}\right)\right\rangle ^{\left(2\right)}\\
 & \times\left\langle a\left(t_{1}\right)\right\rangle ^{\left(1\right)}\\
\left\langle a^{\dagger}\left(t_{2}\right)a^{\dagger}\left(t_{1}\right)a\left(t_{1}\right)a\left(t_{2}\right)\right\rangle ^{\left(4\right)} & =\left\langle a^{\dagger}\left(t_{2}\right)a^{\dagger}\left(t_{1}\right)\right\rangle ^{\left(2\right)}\\
 & \times\left\langle a\left(t_{1}\right)a\left(t_{2}\right)\right\rangle ^{\left(2\right)}
\end{align*}

\section{Calculation of $\left\langle aa\right\rangle ^{\left(2\right)}$\label{a2}}

The system of equations for the same-time 2-operator products in the
second order in $\alpha$ is readily derived from Heisenberg-Langevin
equations. For notational convenience here, we do not explicitly write
superscripts $^{\left(1,2\right)}$, nor the time since we only dealt
with same-time mean values : hence $\left\langle aa\right\rangle $
should be understood as $\left\langle a\left(t\right)a\left(t\right)\right\rangle ^{\left(2\right)}$
and $\left\langle \sigma_{ge}^{\left(i\right)}\right\rangle $ as
$\left\langle \sigma_{ge}^{\left(i\right)}\left(t\right)\right\rangle ^{\left(1\right)}$.
We thus find
\begin{widetext}
\begin{eqnarray*}
\frac{d}{dt}\left\langle aa\right\rangle  & = & 2D_{c}\left\langle aa\right\rangle -2ig\sum_{i}\left\langle a\sigma_{ge}^{\left(i\right)}\right\rangle -2i\alpha\left\langle a\right\rangle \\
\frac{d}{dt}\left\langle a\sigma_{ge}^{\left(i\right)}\right\rangle  & = & \left(D_{c}+D_{e}\right)\left\langle a\sigma_{ge}^{\left(i\right)}\right\rangle -i\frac{\Omega_{b}}{2}\left\langle a\sigma_{gr}^{\left(i\right)}\right\rangle -ig\left\langle aa\right\rangle -ig\sum_{j}\left\langle \sigma_{ge}^{\left(j\right)}\sigma_{ge}^{\left(i\right)}\right\rangle -i\alpha\left\langle \sigma_{ge}^{\left(i\right)}\right\rangle \\
\frac{d}{dt}\left\langle a\sigma_{gr}^{\left(i\right)}\right\rangle  & = & \left(D_{c}+D_{r}\right)\left\langle a\sigma_{gr}^{\left(i\right)}\right\rangle -ig\sum_{j}\left\langle \sigma_{ge}^{\left(j\right)}\sigma_{gr}^{\left(i\right)}\right\rangle -i\alpha\left\langle \sigma_{gr}^{\left(i\right)}\right\rangle -i\frac{\Omega_{b}}{2}\left\langle a\sigma_{ge}^{\left(i\right)}\right\rangle \\
\frac{d}{dt}\left\langle \sigma_{ge}^{\left(j\right)}\sigma_{ge}^{\left(i\right)}\right\rangle  & = & 2D_{e}\left\langle \sigma_{ge}^{\left(j\right)}\sigma_{ge}^{\left(i\right)}\right\rangle -i\frac{\Omega_{b}}{2}\left\langle \sigma_{ge}^{\left(j\right)}\sigma_{gr}^{\left(i\right)}\right\rangle -i\frac{\Omega_{b}}{2}\left\langle \sigma_{gr}^{\left(j\right)}\sigma_{ge}^{\left(i\right)}\right\rangle -ig\left\langle a\sigma_{ge}^{\left(j\right)}\right\rangle -ig\left\langle a\sigma_{ge}^{\left(i\right)}\right\rangle \\
\frac{d}{dt}\left\langle \sigma_{ge}^{\left(j\right)}\sigma_{gr}^{\left(i\right)}\right\rangle  & = & \left(D_{e}+D_{r}\right)\left\langle \sigma_{ge}^{\left(j\right)}\sigma_{gr}^{\left(i\right)}\right\rangle -i\frac{\Omega_{b}}{2}\left\langle \sigma_{gr}^{\left(j\right)}\sigma_{gr}^{\left(i\right)}\right\rangle -ig\left\langle a\sigma_{gr}^{\left(i\right)}\right\rangle -i\frac{\Omega_{b}}{2}\left\langle \sigma_{ge}^{\left(j\right)}\sigma_{ge}^{\left(i\right)}\right\rangle \\
\frac{d}{dt}\left\langle \sigma_{gr}^{\left(j\right)}\sigma_{gr}^{\left(i\right)}\right\rangle  & = & \left(2D_{r}-i\kappa_{i,j}\right)\left\langle \sigma_{gr}^{\left(j\right)}\sigma_{gr}^{\left(i\right)}\right\rangle -i\frac{\Omega_{b}}{2}\left\langle \sigma_{ge}^{\left(j\right)}\sigma_{gr}^{\left(i\right)}\right\rangle -i\frac{\Omega_{b}}{2}\left\langle \sigma_{gr}^{\left(j\right)}\sigma_{ge}^{\left(i\right)}\right\rangle 
\end{eqnarray*}

\end{widetext}

Assuming that the medium is homogeneous, \emph{i.e.} that for all
$\left(i,j\right)$, $\left\langle \sigma_{ge}^{\left(j\right)}\sigma_{gr}^{\left(i\right)}\right\rangle =\left\langle \sigma_{ge}^{\left(i\right)}\sigma_{gr}^{\left(j\right)}\right\rangle $
and $\left\langle a\sigma_{ge}^{\left(i\right)}\right\rangle =\left\langle a\sigma_{ge}^{\left(j\right)}\right\rangle $,
in the steady state this system yields
\begin{widetext}
\begin{eqnarray*}
\left\langle aa\right\rangle  & = & \frac{g}{D_{c}}\sum_{i}\left\langle a\sigma_{ge}^{\left(i\right)}\right\rangle +\frac{\alpha}{D_{c}}\left\langle a\right\rangle \\
\left\langle a\sigma_{ge}^{\left(i\right)}\right\rangle  & = & \frac{\Omega_{b}}{2\left(D_{c}+D_{e}\right)}\left\langle a\sigma_{gr}^{\left(i\right)}\right\rangle +\frac{g}{\left(D_{c}+D_{e}\right)}\left\langle aa\right\rangle +\frac{g}{\left(D_{c}+D_{e}\right)}\sum_{j}\left\langle \sigma_{ge}^{\left(j\right)}\sigma_{ge}^{\left(i\right)}\right\rangle +\frac{\alpha}{\left(D_{c}+D_{e}\right)}\left\langle \sigma_{ge}^{\left(i\right)}\right\rangle \\
\left\langle a\sigma_{gr}^{\left(i\right)}\right\rangle  & = & \frac{g}{\left(D_{c}+D_{r}\right)}\sum_{j}\left\langle \sigma_{ge}^{\left(j\right)}\sigma_{gr}^{\left(i\right)}\right\rangle +\frac{\alpha}{\left(D_{c}+D_{r}\right)}\left\langle \sigma_{gr}^{\left(i\right)}\right\rangle +\frac{\Omega_{b}}{2\left(D_{c}+D_{r}\right)}\left\langle a\sigma_{ge}^{\left(i\right)}\right\rangle \\
\left\langle \sigma_{ge}^{\left(j\right)}\sigma_{ge}^{\left(i\right)}\right\rangle  & = & \frac{\Omega_{b}}{2D_{e}}\left\langle \sigma_{ge}^{\left(j\right)}\sigma_{gr}^{\left(i\right)}\right\rangle +\frac{g}{D_{e}}\left\langle a\sigma_{ge}^{\left(i\right)}\right\rangle \\
\left\langle \sigma_{ge}^{\left(j\right)}\sigma_{gr}^{\left(i\right)}\right\rangle  & = & \frac{\Omega_{b}}{2\left(D_{e}+D_{r}\right)}\left\langle \sigma_{gr}^{\left(j\right)}\sigma_{gr}^{\left(i\right)}\right\rangle +\frac{g}{\left(D_{e}+D_{r}\right)}\left\langle a\sigma_{gr}^{\left(i\right)}\right\rangle +\frac{\Omega_{b}}{2\left(D_{e}+D_{r}\right)}\left\langle \sigma_{ge}^{\left(j\right)}\sigma_{ge}^{\left(i\right)}\right\rangle \\
\left\langle \sigma_{gr}^{\left(j\right)}\sigma_{gr}^{\left(i\right)}\right\rangle  & = & \frac{\Omega_{b}}{2\left(D_{r}-\frac{\kappa_{i,j}}{2}\right)}\left\langle \sigma_{ge}^{\left(j\right)}\sigma_{gr}^{\left(i\right)}\right\rangle 
\end{eqnarray*}

\end{widetext}

Note that the first-order values $\left\langle a\right\rangle \equiv\left\langle a\right\rangle ^{\left(1\right)}$,
$\left\langle \sigma_{ge}^{\left(i\right)}\right\rangle \equiv\left\langle \sigma_{ge}^{\left(i\right)}\right\rangle ^{\left(1\right)}$,
$\left\langle \sigma_{gr}^{\left(i\right)}\right\rangle \equiv\left\langle \sigma_{gr}^{\left(i\right)}\right\rangle ^{\left(1\right)}$
have been determined through solving the first-order steady state
system, see Eqs. (\ref{a1}-\ref{sigmagr1}) in the main text.

Summing the above equations over atom numbers $\left(i,j\right)$
yields a system on averages of the collective operators $b\equiv\frac{1}{\sqrt{N}}\sum_{i}\sigma_{ge}^{\left(i\right)}$
and $c\equiv\frac{1}{\sqrt{N}}\sum_{i}\sigma_{gr}^{\left(i\right)}$
and field operator $a$, which is\emph{ almost} closed but for the
last equation which will now be considered and approximated. Eliminating
$\left\langle \sigma_{ge}^{\left(j\right)}\sigma_{gr}^{\left(i\right)}\right\rangle $
and $\left\langle \sigma_{ge}^{\left(j\right)}\sigma_{ge}^{\left(i\right)}\right\rangle $
from the last three equations we get 
\begin{eqnarray*}
 &  & \left\langle \sigma_{gr}^{\left(j\right)}\sigma_{gr}^{\left(i\right)}\right\rangle =\\
 &  & \frac{\Omega_{b}g}{2\left\{ \left(D_{r}-\frac{\kappa_{i,j}}{2}\right)\left[\left(D_{e}+D_{r}\right)-\frac{\Omega_{b}^{2}}{4D_{e}}\right]-\frac{\Omega_{b}^{2}}{4}\right\} }\left\langle a\sigma_{gr}^{\left(i\right)}\right\rangle \\
 &  & +\frac{\Omega_{b}^{2}g}{4D_{e}\left\{ \left(D_{r}-\frac{\kappa_{i,j}}{2}\right)\left[\left(D_{e}+D_{r}\right)-\frac{\Omega_{b}^{2}}{4D_{e}}\right]-\frac{\Omega_{b}^{2}}{4}\right\} }\left\langle a\sigma_{ge}^{\left(i\right)}\right\rangle 
\end{eqnarray*}
We now sum over $i$ and $j$ indices and divide by $N$ this equation
to get

\begin{eqnarray*}
\left\langle cc\right\rangle  & = & \frac{\Omega_{b}g}{2}\sum_{i}K_{i}\left\langle a\sigma_{gr}^{\left(i\right)}\right\rangle +\frac{\Omega_{b}^{2}g}{4D_{e}}\sum_{i}K_{i}\left\langle a\sigma_{ge}^{\left(i\right)}\right\rangle 
\end{eqnarray*}
where we introduced the coefficient 
\[
K_{i}\equiv\frac{1}{N}\sum_{j}\frac{1}{\left(D_{e}+D_{r}-\frac{\Omega_{b}^{2}}{4D_{e}}\right)\left(D_{r}-\frac{\kappa_{i,j}}{2}\right)-\frac{\Omega_{b}^{2}}{4}}.
\]
Making the approximation that $K_{i}$ does not depend on $i$, \emph{i.e.}
$K_{i}\approx K$, we get:

\[
\left\langle cc\right\rangle \approx\frac{\Omega_{b}g\sqrt{N}}{2}K\left\langle ac\right\rangle +\frac{\Omega_{b}^{2}g\sqrt{N}}{4D_{e}}K\left\langle ab\right\rangle 
\]
To estimate $K$ we consider that the sample is a sphere of radius\textbf{
$R$}

\begin{eqnarray*}
K & = & \frac{1}{N}\sum_{j}\frac{1}{\left(D_{e}+D_{r}-\frac{\Omega_{b}^{2}}{4D_{e}}\right)\left(D_{r}-\frac{\kappa_{i,j}}{2}\right)-\frac{\Omega_{b}^{2}}{4}}\\
 & \approx & \frac{4\pi}{\frac{4\pi}{3}R^{3}}\int_{0}^{R}\frac{r^{2}}{\left(D_{e}+D_{r}-\frac{\Omega_{b}^{2}}{4D_{e}}\right)\left(D_{r}-\frac{C_{6}}{2r^{6}}\right)-\frac{\Omega_{b}^{2}}{4}}dr\\
 & = & \frac{3}{R^{3}}\int_{0}^{R}\frac{r^{2}}{\left(D_{e}+D_{r}-\frac{\Omega_{b}^{2}}{4D_{e}}\right)\left(D_{r}-\frac{C_{6}}{2r^{6}}\right)-\frac{\Omega_{b}^{2}}{4}}dr
\end{eqnarray*}
For large values of $R$, $K$ does not depend on the geometry 
\begin{align*}
K\underset{R\rightarrow\infty}{\sim}\frac{1}{\left(D_{e}+D_{r}-\frac{\Omega_{b}^{2}}{4D_{e}}\right)D_{r}-\frac{\Omega_{b}^{2}}{4}}\\
\times\left(1-\frac{\sqrt{2}\pi^{2}}{3V}\sqrt{\frac{C_{6}}{\frac{\Omega_{b}^{2}}{4\left(D_{e}+D_{r}-\frac{\Omega_{b}^{2}}{4D_{e}}\right)}-D_{r}}}\right)
\end{align*}

Finally the desired closed system is
\begin{widetext}
\begin{eqnarray}
\left\langle aa\right\rangle  & = & \frac{g\sqrt{N}}{D_{c}}\left\langle ab\right\rangle +\frac{\alpha}{D_{c}}\left\langle a\right\rangle \nonumber \\
\left\langle ab\right\rangle  & = & \frac{\Omega_{b}}{2\left(D_{c}+D_{e}\right)}\left\langle ac\right\rangle +\frac{g\sqrt{N}}{\left(D_{c}+D_{e}\right)}\left\langle aa\right\rangle +\frac{g\sqrt{N}}{\left(D_{c}+D_{e}\right)}\left\langle bb\right\rangle +\frac{\alpha}{\left(D_{c}+D_{e}\right)}\left\langle b\right\rangle \nonumber \\
\left\langle ac\right\rangle  & = & \frac{g\sqrt{N}}{\left(D_{c}+D_{r}\right)}\left\langle bc\right\rangle +\frac{\alpha}{\left(D_{c}+D_{r}\right)}\left\langle c\right\rangle +\frac{\Omega_{b}}{2\left(D_{c}+D_{r}\right)}\left\langle ab\right\rangle \\
\left\langle bb\right\rangle  & = & \frac{\Omega_{b}}{2D_{e}}\left\langle bc\right\rangle +\frac{g\sqrt{N}}{D_{e}}\left\langle ab\right\rangle \nonumber \\
\left\langle bc\right\rangle  & = & \frac{\Omega_{b}}{2\left(D_{e}+D_{r}\right)}\left\langle cc\right\rangle +\frac{g\sqrt{N}}{\left(D_{e}+D_{r}\right)}\left\langle ac\right\rangle +\frac{\Omega_{b}}{2\left(D_{e}+D_{r}\right)}\left\langle bb\right\rangle \nonumber \\
\left\langle cc\right\rangle  & = & \frac{\Omega_{b}g\sqrt{N}}{2}K\left\langle ac\right\rangle +\frac{\Omega_{b}^{2}g\sqrt{N}}{4D_{e}}K\left\langle ab\right\rangle \nonumber 
\end{eqnarray}

\end{widetext}

which allows to determine $\left\langle aa\right\rangle $. The analytical
solution is too cumbersome to be displayed in this paper but can be
readily obtained by matrix inversion.

\section{Factorization in the presence of extra dephasing\label{FactDeph}}

In this appendix, we show in which conditions the factorization of
field operator products described in Appendix \ref{Factorization}
remains valid in the presence of extra dephasing due to laser frequency
and intensity noise. Such dephasing is correctly accounted for by
adding the term $-\gamma_{d}\sigma_{gr}^{\left(n\right)}+F_{gr}^{\left(d\right)}$
in the Heisenberg-Langevin equation Eq. (\ref{HL3}) on $\sigma_{gr}^{\left(n\right)}$,
where $F_{gr}^{\left(d\right)}$ is an extra Langevin force and $\gamma_{d}\approx0.15\times\gamma_{e}$,
$\gamma_{r}\approx0.01\times\gamma_{e}$ and $\gamma_{e}=2\pi\times3$
MHz in the experimental setup.

In the absence of interatomic interactions, because laser and cavity
fields address the atoms symmetrically, the ensemble evolves in the
subspace of symmetric states. The atomic system essentially remains
in this subspace, even when the interactions are taken into account,
if the number of Rydberg excitations in the sample is much less than
the total number of Rydberg bubbles the ensemble can accomodate for.
Such symmetric superpositions actually not only contain ``allowed''
components (\emph{i.e.} with Rydberg atoms further than a Rydberg
bubble radius apart) but also ``forbidden'' components (with Rydberg
atoms closer than a Rydberg bubble radius). Their number is, however,
very small compared to that of ``allowed'' confgurations and they
will therefore only slightly alter the outcome of dissipative dynamics
of the system.

Under these assumptions, let us show in which conditions the mean
value $\left\langle c^{\dagger}c\right\rangle $ factorizes at lowest
order. Focusing on the dissipative part of Bloch equations for $\sigma_{gr}^{\left(i\right)}$
and $\sigma_{rr}^{\left(i\right)}$ (note that for the latter, there
is no extra dephasing) we get 
\begin{align*}
\frac{d}{dt}\left.\left\langle \sigma_{gr}^{\left(i\right)}\right\rangle \right|_{d} & =-\left(\gamma_{r}+\gamma_{d}\right)\left\langle \sigma_{gr}^{\left(i\right)}\right\rangle \\
\frac{d}{dt}\left.\left\langle \sigma_{rg}^{\left(i\right)}\sigma_{gr}^{\left(j\right)}\right\rangle \right|_{d,i\neq j} & =-2\left(\gamma_{r}+\gamma_{d}\right)\left\langle \sigma_{rg}^{\left(i\right)}\sigma_{gr}^{\left(j\right)}\right\rangle \\
\frac{d}{dt}\left.\left\langle \sigma_{rr}^{\left(i\right)}\right\rangle \right|_{d} & =-2\gamma_{r}\left\langle \sigma_{rr}^{\left(i\right)}\right\rangle 
\end{align*}
and recalling that $c\equiv\frac{1}{\sqrt{N}}\sum_{i}\sigma_{gr}^{\left(i\right)}$,
we get $\left\langle c^{\dagger}c\right\rangle =\frac{1}{N}\sum_{i}\left\langle \sigma_{rr}^{\left(i\right)}\right\rangle +\frac{1}{N}\sum_{i\neq j}\left\langle \sigma_{rg}^{\left(i\right)}\sigma_{gr}^{\left(j\right)}\right\rangle $
whence, for a short time interval
\begin{align*}
\left.\frac{d\langle c^{\dagger}c\rangle}{dt}\right|_{d} & =\frac{1}{N}\sum_{i}\frac{d}{dt}\left.\langle\sigma_{rr}^{\left(i\right)}\rangle\right|_{d}+\frac{1}{N}\sum_{i\neq j}\frac{d}{dt}\left.\langle\sigma_{rg}^{\left(i\right)}\sigma_{gr}^{\left(j\right)}\rangle\right|_{d}\\
 & =-\frac{2\gamma_{r}}{N}\sum_{i}\langle\sigma_{rr}^{\left(i\right)}\rangle-\frac{2}{N}\left(\gamma_{r}+\gamma_{d}\right)\sum_{i\neq j}\langle\sigma_{rg}^{\left(i\right)}\sigma_{gr}^{\left(j\right)}\rangle\\
 & =-\frac{2\gamma_{r}}{N}\sum_{i}\langle\sigma_{rr}^{\left(i\right)}\rangle+\frac{2}{N}\left(\gamma_{r}+\gamma_{d}\right)\sum_{i}\langle\sigma_{rr}^{\left(i\right)}\rangle+\\
 & -\frac{2}{N}\left(\gamma_{r}+\gamma_{d}\right)\sum_{i,j}\langle\sigma_{rg}^{\left(i\right)}\sigma_{gr}^{\left(j\right)}\rangle\\
 & =\frac{2\gamma_{d}}{N}\sum_{i}\langle\sigma_{rr}^{\left(i\right)}\rangle-\frac{2}{N}\left(\gamma_{r}+\gamma_{d}\right)\sum_{i,j}\langle\sigma_{rg}^{\left(i\right)}\sigma_{gr}^{\left(j\right)}\rangle\\
\frac{d}{dt}\left.\langle c^{\dagger}c\rangle\right|_{d} & =\frac{2\gamma_{d}}{N}\sum_{i}\langle\sigma_{rr}^{\left(i\right)}\rangle-2\left(\gamma_{r}+\gamma_{d}\right)\langle c^{\dagger}c\rangle
\end{align*}

When there are $n_{r}$ Rydberg excitations in the sample, with $n_{r}\ll N_{b}\ll N$
($N_{b}$ is the maximum number of Rydberg excitations the sample
can contain), one has $\left\langle c^{\dagger}c\right\rangle \approx\sum_{i}\left\langle \sigma_{rr}^{\left(i\right)}\right\rangle \approx n_{r}$
whence 
\[
\frac{d}{dt}\left.\left\langle c^{\dagger}c\right\rangle \right|_{d}\approx-2\left[\gamma_{r}+\gamma_{d}\left(1-\frac{1}{N}\right)\right]\left\langle c^{\dagger}c\right\rangle 
\]
and for $\gamma_{r}\ll\gamma_{d}\ll N\gamma_{r}$ 
\[
\frac{d}{dt}\left.\left\langle c^{\dagger}c\right\rangle \right|_{d}\approx-2\gamma_{d}\left\langle c^{\dagger}c\right\rangle 
\]
so, from the point of view of $c^{\dagger}c$, everything works as
if the system was radiatively damped with the rate $\gamma_{d}$.
In the same conditions, we moreover have 
\[
\frac{d}{dt}\left.\left\langle c\right\rangle \right|_{d}\approx-\gamma_{d}\left\langle c\right\rangle 
\]
and again, from the point of view of $c$, everything works as if
the system was radiatively damped with the rate $\gamma_{d}$. Moreover,
since all other dynamical equations (for population, coherence and
field operator mean values) remain formally the same as in the purely
radiative damping, the factorization procedure remains valid for $\left\langle a^{\dagger}a\right\rangle $
provided that $\gamma_{r}\ll\gamma_{d}\ll N\gamma_{r}$ and the radiative
coherence decay $\gamma_{r}$ is effectively replaced by the dephasing
decay rate $\gamma_{d}$.

This result can also be extended to higher order quantities $\left\langle \left(a^{\dagger}\right)^{m}a^{p}\right\rangle $.

\end{document}